\documentclass[twocolumn,showpacs,preprintnumbers,amsmath,amssymb]{revtex4}

\usepackage{graphicx}
\usepackage{dcolumn}
\usepackage{bm}

\begin{document}


\title{Polyelectrolyte Layer-by-Layer Assembly from Self-Consistent Field Calculations}

\author{Qiang Wang}
\email{q.wang@colostate.edu}
\affiliation{Department of Chemical Engineering, Colorado State University, Fort Collins, CO 80523-1370}

\date{\today}

\begin{abstract}
We have modeled the layer-by-layer assembly process of flexible polyelectrolytes on flat surfaces. The multilayer has a three-zone structure. An exponential growth is found for the first several layers, followed by a linear growth for subsequent layers evolving towards a steady state. While adjacent layers are highly interpenetrating, stratification can be seen for every four or more layers. The effects of surface charge density, bulk salt concentration, and solvent quality on the thickness and internal structure of the multilayer are also studied. Our results agree with experimental findings.
\end{abstract}

\maketitle

Proposed in early 90's, the polyelectrolyte (PE) layer-by-layer assembly technique has attracted exponentially increasing interest due to its simplicity, versatility, and potential applications.\cite{Decher,BerRev,SchRev} Through alternating exposure of a charged substrate to solutions of polyanions and polycations, hundreds of thin adsorbed layers of polyions can be readily built-up on the substrate. In addition to charged polymers, charged nano-particles and platelets have also been used.\cite{BerRev} This technique has a wide variety of potential applications, including surface modification, sensors, conducting or light-emitting devices, drug delivery, nano-reactors, etc.\cite{BerRev}

Our understanding on the formation mechanism, internal structure, and molecular property of PE multilayer, however, is still at an early stage.\cite{SchRev} In great contrast to thousands of experimental papers on the layer-by-layer assembly, very few theoretical\cite{Kotov,JoanM2} or simulation\cite{Mess,DobrS,DobrM} studies have been reported. In general, the build-up of PE multilayers is based on the charge inversion, i.e., the charge of newly adsorbed PE inverts (overcompensates) the existing charge in the film (including the bare substrate charge). Using a self-consistent field (SCF) theory\cite{WQ1}, we recently examined in detail the effects of various parameters in the adsorption of flexible PE on flat surfaces; strong charge inversion was found for relatively long PE on oppositely charged, attractive surfaces in poor solvent at high salt concentrations.\cite{WQ2} In this work we use the same formalism to model the process of PE layer-by-layer assembly and to examine the thickness and internal structure of the multilayer.

We only consider monovalent systems, and assume that all ions carrying the same type of charge are identical. We further ignore the volume of small ions, and impose the incompressibility constraint everywhere in the system. Under the ground-state dominance approximation\cite{GSDA}, the SCF equations describing the adsorption of charged polymer A, in the presence of polymers 1 and 2 (of which density profiles are fixed), from its bulk solution of small-molecule solvent S (having the same density $\rho_0$ as all the polymer segments) onto a flat surface are given by\cite{WQ2}
{\setlength \arraycolsep{1pt}
\begin{eqnarray} \label{eq:MFE}
\frac{\mathrm{d}^2}{\mathrm{d} x^2} \sqrt{\phi_\mathrm{A}(x)} & = & \Big[ \omega_\mathrm{A}(x) - N g_\mathrm{A}(x) - \omega_0 \Big] \sqrt{\phi_\mathrm{A}(x)} \\
\frac{\epsilon}{N} \frac{\mathrm{d}^2 \psi(x)}{\mathrm{d} x^2} & = & \sum_\mathrm{P} \phi_\mathrm{P}(x) \frac{\mathrm{d}g_\mathrm{P}(x)}{\mathrm{d}\psi(x)} + v_\mathrm{A} p_\mathrm{A} \phi_{\mathrm{A},b} \exp\Big[v_\mathrm{A} \psi(x)\Big] \nonumber \\
& & + 2 c_{s,b} \sinh \psi(x)
\end{eqnarray} }
where $x$ denotes the direction perpendicular to the surface (placed at $x=0$); $\phi_\mathrm{P}(x)$ is the normalized (by $\rho_0$) segmental density of polymer P (= A, 1, and 2); $\psi(x)$ is the electrostatic potential (in units of $k_B T / e$, where $k_B$ is the Boltzmann constant, $T$ the absolute temperature, and $e$ the elementary charge), and we set $k_B T = 1$ hereafter; $\omega_\mathrm{A}(x) = N \{ \sum_\mathrm{P} (\chi_{\mathrm{PA}}-\chi_{\mathrm{PS}}) \phi_\mathrm{P}(x) + \chi_\mathrm{AS} [1 - \sum_\mathrm{P} \phi_\mathrm{P}(x)] - \ln [1 - \sum_\mathrm{P} \phi_\mathrm{P}(x)] \}$, where $\chi_\mathrm{PP'}$ ($\mathrm{P'}$ = A, 1, and 2) and $\chi_\mathrm{PS}$ are the Flory-Huggins parameters ($\chi_\mathrm{PP}=0$); $g_\mathrm{P}(x) \equiv - v_\mathrm{P} p_\mathrm{P} \psi(x)$ for the smeared charge distribution where each segment of polymer P carries charge $v_\mathrm{P} p_\mathrm{P} e$ (a model for strongly dissociating PE), and $g_\mathrm{P}(x) \equiv \ln [1-p_\mathrm{P}+p_\mathrm{P}\exp(-v_\mathrm{P}\psi(x))]$ for the annealed charge distribution where each segment of P has a probability $p_\mathrm{P}$ of carrying charge $v_\mathrm{P} e$ (a model for weakly dissociating PE), and we set $v_1 = -v_2 = -1$; $\phi_{\mathrm{A},b}$ and $c_{s,b}$ are the normalized (by $\rho_0$) segmental density of A and salt concentration (we assume that one salt molecule generates one cation), respectively, in the bulk solution placed at $x=l>0$. Note that we have normalized the distance by $R_g \equiv a\sqrt{N/6}$, where $a$ is the statistical segment length of polymers (assumed to be the same for all P), and $N$ is an (arbitrary) chain length chosen for the normalization; we have also normalized the dielectric constant $\epsilon$ (assumed to be uniform for $x \ge 0$) by $2 \pi \rho_0 e^2 a^2 / (3 k_B T)$, where $\epsilon$ before the normalization is in units of $4 \pi \epsilon_0$ with $\epsilon_0=8.854 \times 10^{-12}$ (A$\cdot$s)$^2$/(J$\cdot$m) being the permittivity of vacuum. Note further that we have chosen $\psi(l)=0$ and $\phi_\mathrm{A}(l) = \phi_{\mathrm{A},b}$, which effectively couple the bulk solution to our system for both ions and polymer A, and thus give $\omega_0 = \omega_\mathrm{A}(l)$; they are satisfied by choosing large enough system size $l$ such that both $|\phi_\mathrm{A}(l)-\phi_{\mathrm{A},b}|$ and $|\psi(l)|$ are less than $10^{-10}$.

The boundary conditions for the mean-field equations are\cite{WQ2}: $(\mathrm{d} \ln \sqrt{\phi_\mathrm{A}} / \mathrm{d} x) |_{x=0} = d^{-1}$, where $d^{-1}$ represents the short-range (non-Coulombic) surface-polymer interactions and includes both energetic and entropic contributions; $(\mathrm{d} \sqrt{\phi_\mathrm{A}} / \mathrm{d}x ) |_{x=l} = 0$; $(\mathrm{d}\psi/\mathrm{d}x)|_{x=0} = - \sigma_\mathrm{SF}\sqrt{N}/\epsilon$, where the normalized surface charge density $\sigma_\mathrm{SF} \equiv \sigma_0 \sqrt{6} / (\rho_0 a) > 0$, and $\sigma_0$ is the number of surface charges (in units of $e$) per unit area; and $(\mathrm{d}\psi/\mathrm{d}x)|_{x=l} = 0$, which ensures the overall charge neutrality of our system (including the surface charge). The mean-field equations are solved by a relaxation method\cite{Relax}.

We do the following to mimic the process of layer-by-layer assembly of polymers 1 (polyanions) and 2 (polycations): (1) For the $i^\mathrm{th}$ deposition, A is set to represent polymer 1 when $i$ is odd and 2 when $i$ is even. After solving the mean-field equations, the amount of adsorbed polymers in this deposition, $\Gamma^{(i)} \equiv \sqrt{N} \int_0^l [\phi_\mathrm{A}^{(i)}(x)-\phi_{\mathrm{A},b}] \mathrm{d}x$, is calculated. (2) A washing step follows each deposition and modifies the polymer density profile from $\phi_\mathrm{A}^{(i)}(x)$ to $\phi_{\mathrm{A},w}^{(i)}(x)$. Since it is not possible to distinguish between adsorbed and unabsorbed chains under the ground-state dominance approximation, we (somewhat arbitrarily) choose $\phi_{\mathrm{A},w}^{(i)}(x) = \phi_\mathrm{A}^{(i)}(x) \exp[t x^2/(x-x_w^{(i)})]$ for $x<x_w^{(i)}$ and $\phi_{\mathrm{A},w}^{(i)}(x) = 0$ for $x \ge x_w^{(i)}$, where the determination of the ``washing point'' $x_w^{(i)} \in (0,l]$ is explained in Ref.~\cite{Note2}, and $t>0$ is calculated according to $\sqrt{N} \int_0^{x_w^{(i)}} \phi_{\mathrm{A},w}^{(i)}(x) \mathrm{d}x = \Gamma^{(i)}$. (3) $\phi_1(x)$ and $\phi_2(x)$ to be used in the $i+1^\mathrm{th}$ deposition are accumulated for polyanions and polycations, respectively, from $\phi_{\mathrm{A},w}^{(j)}(x)$ for all $j \le i$. That in each deposition $\phi_1(x)$ and $\phi_2(x)$ are fixed represents the irreversibility of PE adsorption\cite{SchRev}; we therefore model the multilayer growth using a series of kinetically trapped states.

\begin{figure}[t]
\begin{center}
\includegraphics[height=5.5cm]{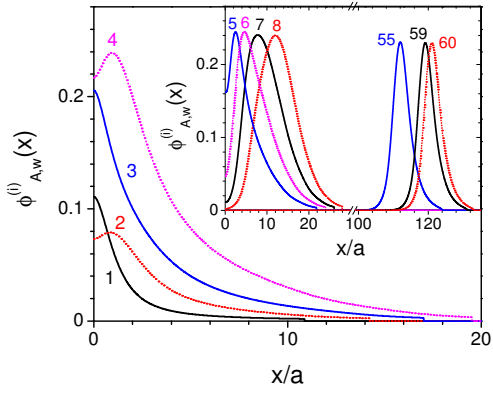}
\end{center}
\caption{\label{phiAw}(Color online) Polymer segmental density profile of the $i^\mathrm{th}$ layer $\phi_{\mathrm{A},w}^{(i)}(x)$. The layer number $i$ is marked near the corresponding profile. $\sigma_\mathrm{SF}=0.1$, $c_{s,b}=0.05$, and $\chi_{12} = 0$.}
\end{figure}

\begin{figure}[t]
\begin{center}
\includegraphics[height=5.5cm]{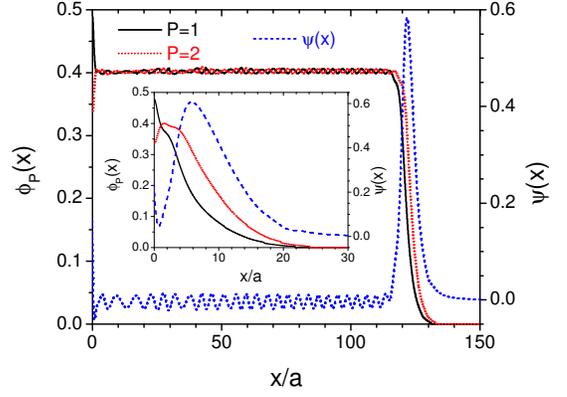}
\end{center}
\caption{\label{phiAC_psi}(Color online) Total segmental density profiles of polyanions and polycations after washing, $\phi_1(x)$ and $\phi_2(x)$, respectively, and the electrostatic potential profile before washing, $\psi(x)$, after the $60^\mathrm{th}$ deposition. The inset shows the same profiles after the $6^\mathrm{th}$ deposition. $\sigma_\mathrm{SF}=0.1$, $c_{s,b}=0.05$, and $\chi_{12} = 0$.}
\end{figure}

In this Letter we study the multilayer formation on an indifferent surface ($d^{-1}=0$) using symmetric PE ($p_\mathrm{P} = 0.5$ and $\phi_{\mathrm{A},b}=7.5 \times 10^{-4}$) both having smeared charge distribution in a poor solvent ($\chi_\mathrm{PS} = 1$)\cite{Note3}. Fig.~\ref{phiAw} shows the polymer segmental density profile of the $i^\mathrm{th}$ deposition after washing (referred to as the $i^\mathrm{th}$ layer), $\phi_{\mathrm{A},w}^{(i)}(x)$, on a positively charged substrate of $\sigma_\mathrm{SF}=0.1$, where a bulk salt concentration of $c_{s,b}=0.05$ is used in each deposition.\cite{Note1} Under such conditions, the first layer strongly overcompensates the substrate charge: $\Sigma \sigma^{(1)} / \sigma_\mathrm{SF} \approx -1.36$, where $\Sigma \sigma^{(i)} \equiv \sigma_\mathrm{SF} + \sum_{j=1}^i v_\mathrm{A}^{(j)} p_\mathrm{A} \Gamma^{(j)}$ is the overall charge of the multilayer (including the substrate charge) after the $i^\mathrm{th}$ deposition. We set $\chi_{12}=0$ in this case; it is therefore the presence of the first layer of polyanions that leads to the adsorption of the second layer (i.e., $\Gamma^{(2)} > 0$), which carries the same type of charge as the substrate. Similarly, we have $\Gamma^{(i)} > \Gamma^{(i-2)}$ for $i=3$ to 8. Subsequent layers have similar density profiles to $\phi_{\mathrm{A},w}^{(8)}(x)$, and are further away from the substrate; consistent with experimental observations\cite{Decher,BerRev,SchRev}, Fig.~\ref{phiAw} also shows that, while adjacent layers are highly interpenetrating, stratification can be seen for every four or more layers.

The total segmental density profiles of polyanions and polycations, $\phi_1(x)$ and $\phi_2(x)$, respectively, after the $60^\mathrm{th}$ deposition and washing are shown in Fig.~\ref{phiAC_psi}; also shown is the electrostatic potential profile $\psi(x)$ after the $60^\mathrm{th}$ deposition (before washing). Three zones can be identified in the multilayer: In the interior of the film (Zone II), $\phi_1(x)$ and $\phi_2(x)$ are almost constant and $\psi(x) \approx 0$. Here, $\phi_1(x) \approx \phi_2(x)$ due to the symmetric PE we used, and the concentrations of ions (before washing) are close to those in the bulk solution. We see small oscillations in $\phi_1(x)$ and $\phi_2(x)$ (which are out-of-phase), indicating the weakly spatial enrichment of polyanions and polycations; $\psi(x)$ also oscillates, in-phase with $\phi_2(x)$, in Zone II. Close to the substrate (Zone I), $\phi_1(x) > \phi_2(x)$ due to the electrostatic interactions between the substrate and polymers. In the immediate vicinity of the substrate, $\psi(x)$ is positive and larger than that in Zone II. At the outer boundary of the multilayer (Zone III), both $\phi_1(x)$ and $\phi_2(x)$ decay to 0, and $\psi(x)$ exhibits a large peak (positive when $i$ is even and negative when $i$ is odd). Such a three-zone structure was first proposed by Castelnovo and Joanny for multilayers of flexible and symmetric PE on an indifferent surface in $\theta$-solvent at high salt concentrations, where the polymer density in Zone II was assumed to be constant and equal to that of neutral complex formed in the bulk.\cite{JoanM2} It is also supported by the molecular dynamics simulations of Patel \emph{et~al.}, who studied the process of layer-by-layer assembly of symmetric PE on a planar substrate in a poor solvent with no added salt; the density oscillations in Zone II were also observed by these authors.\cite{DobrM}

\begin{table}[t]
\caption{\label{phiPEM} Average total polymer segmental density in Zone II of the multilayer $\phi_\mathrm{PEM}$. Since the total polymer segmental density slightly varies in Zone II, we use three times its standard deviation as the error. $\sigma_\mathrm{SF}=0.1$ and $c_{s,b}=0.05$. Refer to text for more details.}
\begin{ruledtabular}
\begin{tabular}{ccccc}
$\chi_{12}$ & $\phi_\mathrm{PEM}$ & $\phi_\mathrm{min,N}$ & $\phi_\mathrm{cmp}$ & $\phi_\mathrm{min}$ \\
\hline
0   & $0.8046 \pm 0.0044$ & 0.7968 & 0.6838 & 0.8444 \\
0.5 & $0.7199 \pm 0.0050$ & 0.7127 & 0.5971 & 0.7956 \\
1   & $0.5907 \pm 0.0079$ & 0.5828 & 0.4743 & 0.7354 \\
\end{tabular}
\end{ruledtabular}
\end{table}

\begin{figure}[t]
\begin{center}
\includegraphics[height=5.5cm]{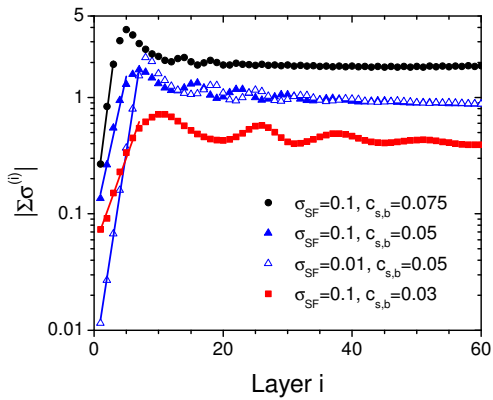}
\end{center}
\caption{\label{Ssgm}(Color online) Semi-logarithmic plot of the overall charge of the multilayer $\Sigma \sigma^{(i)}$ as a function of layer number $i$ ($\Sigma \sigma^{(i)}<0$ when $i$ is odd and $\Sigma \sigma^{(i)}>0$ when $i$ is even). The straight lines are fits for the first several layers exhibiting an exponential growth. $\chi_{12} = 0$.}
\end{figure}

For a homogeneous system, the SCF theory reduces to the Flory-Huggins theory modified for PE\cite{Warn}; in our case of symmetric PE 1 and 2 (with chain lengths $N_\mathrm{P} \to \infty$) having smeared charge distribution in solvent S, the free energy (of mixing) per polymer segment or solvent molecule becomes
{\setlength \arraycolsep{1pt}
\begin{eqnarray} \label{eq:fFH}
f & = & \frac{\phi}{N_\mathrm{P}} \ln \frac{\phi}{2} + (1-\phi) \ln (1-\phi) + \chi_\mathrm{PS} \phi (1-\phi) + \chi_{12} \frac{\phi^2}{4} \nonumber \\
& & + (p_\mathrm{P} \phi + 2 c_s) \ln \frac{p_\mathrm{P} \phi + 2 c_s}{2}
\end{eqnarray} }
where $\phi$ ($= 2 \phi_1 = 2 \phi_2$) and $c_s$ are the total polymer segmental density and salt concentration in the homogeneous system, respectively. From this, the binodal curve describing the phase equilibrium between a polymer-rich phase (where the total polymer segmental density is denoted by $\phi_\mathrm{cmp}$) and a polymer-poor phase (where the polymer density is 0) can be derived, which is given by $\ln (1-\phi_\mathrm{cmp}) / \phi_\mathrm{cmp} + (\chi_\mathrm{PS} - \chi_{12}/4) \phi_\mathrm{cmp} + 1 = 0$. We also define $\phi_\mathrm{min}$ as the total polymer segmental density that minimizes $f$ (with $c_s = c_{s,b}$), and $\phi_\mathrm{min,N}$ as the one that minimizes the neutral part of $f$, i.e., the first line of Eq.~\ref{eq:fFH}. Table~\ref{phiPEM} shows that the average total polymer segmental density in Zone II of the multilayer, denoted by $\phi_\mathrm{PEM}$, is very close to $\phi_\mathrm{min,N}$, instead of $\phi_\mathrm{cmp}$; our results also show that $\phi_\mathrm{PEM}$ is independent of $c_{s,b}$ and $\sigma_\mathrm{SF}$.

\begin{figure}[t]
\begin{center}
\includegraphics[height=5.5cm]{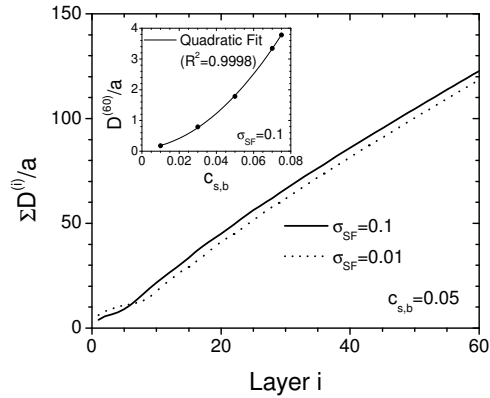}
\end{center}
\caption{\label{SD1} Overall thickness of the multilayer $\Sigma D^{(i)}$ as a function of layer number $i$. The inset shows the thickness of the $60^\mathrm{th}$ layer $D^{(60)}$ as a function of bulk salt concentration $c_{s,b}$. $\chi_{12} = 0$.}
\end{figure}

The inset of Fig.~\ref{phiAC_psi} shows that Zone I is almost fully developed after the first six layers; subsequent layers start to move away from the substrate (as shown in Fig.~\ref{phiAw}). The first six layers exhibit an (approximately) exponential growth, different from the subsequent layers evolving towards a steady state in a damped oscillatory manner. This is shown in Fig.~\ref{Ssgm}, where $\Sigma \sigma^{(i)}<0$ when $i$ is odd and $\Sigma \sigma^{(i)}>0$ when $i$ is even; each layer therefore inverts the overall charge of the multilayer. An almost constant $|\Sigma \sigma^{(i)}|$ is reached in the steady state, which implies that the charge of the newly adsorbed layer $i$ fully inverts the overall charge of the existing multilayer, i.e., $v_\mathrm{A}^{(i)} p_\mathrm{A} \Gamma^{(i)} / \Sigma \sigma^{(i-1)} \approx -2$; this was also found by Dobrynin and co-workers\cite{DobrS,DobrM}. Fig.~\ref{SD1} shows how the overall thickness of the multilayer, $\Sigma D^{(i)} \equiv 2 R_g \int_0^l x \sum_{j=1}^i \phi_{\mathrm{A},w}^{(j)}(x) \mathrm{d}x / \sum_{j=1}^i \Gamma^{(j)}$, evolves with the layer number $i$; an (approximately) linear growth is seen after the first six layers. These are in good agreement with experimental observations\cite{BerRev,SchRev}.

The above characteristics are seen in all of our calculations when multilayer forms. The effects of $\sigma_\mathrm{SF}$ and $c_{s,b}$ are also shown in Figs.~\ref{Ssgm} and \ref{SD1}. The surface charge density $\sigma_\mathrm{SF}$ affects only the exponential growth of the multilayer (and thus $\Sigma D^{(i)}$), but not the adsorbed amount $\Gamma^{(i)}$ (or equivalently $\Sigma \sigma^{(i)}$) and layer thickness in the steady state. The bulk salt concentration $c_{s,b}$ plays a critical role in determining $\Gamma^{(i)}$ and the layer thickness (and thus $\Sigma D^{(i)}$); the inset of Fig.~\ref{SD1} shows a quadratic relationship between the thickness of the $60^\mathrm{th}$ layer (where a linear growth is reached in all cases), $D^{(60)} \equiv \Sigma D^{(60)} - \Sigma D^{(59)}$, and $c_{s,b}$. Note that our mean-field theory does not give strong charge inversion\cite{WQ2} and multilayer formation at low salt concentrations, since the fluctuations and lateral correlations in the system are ignored.

Finally, we also used $\theta$-solvent where $\chi_\mathrm{PS}=0.5$ (with $\sigma_\mathrm{SF}=0.1$ and $\chi_{12} = 0$); no multilayer formation is seen, even at $c_{s,b}=0.1$ where the charge inversion of the first layer ($\Sigma \sigma^{(1)} / \sigma_\mathrm{SF} \approx -0.46$) is stronger than that in the poor solvent at $c_{s,b}=0.01$ (where $\Sigma \sigma^{(1)} / \sigma_\mathrm{SF} \approx -0.27$ and multilayer forms). As noted above, $\Gamma^{(i)} > \Gamma^{(i-2)}$ is observed during the exponential growth of the multilayer in the poor solvent (except for $i=3$ in the $c_{s,b}=0.01$ case). The opposite, however, occurs in the $\theta$-solvent, where $\Gamma^{(i)}$ monotonically decays towards 0. The solvent quality therefore also plays an important role in the multilayer formation. This agrees with the molecular dynamics simulations of Dobrynin and co-workers, who found that multilayer forms under poor solvent conditions but not under good solvent conditions\cite{DobrS,DobrM}.

Financial support for this work was provided by Colorado State University.

\bibliographystyle{unsrt}

\end{document}